\documentstyle[12pt,epsfig]{article}
\newlength{\abstractwidth}
\begin{document}
\thispagestyle{empty}
\pagestyle{plain}
\input{epsf}
\def\beq{\begin{eqnarray}}
\def\eeq{\end{eqnarray}}
\def\nn{\nonumber\\}
\renewcommand{\thefootnote}{\fnsymbol{footnote}}
\renewcommand{\thanks}[1]{\footnote{#1}} 
\newcommand{\starttext}{
\setcounter{footnote}{0}
\renewcommand{\thefootnote}{\arabic{footnote}}}
\def\gb{\bar g}
\def\htt{{\tilde h}}
\def\hb{\bar h}
\def\hbt{{\tilde {\bar h}}}
\def\gt{{\tilde g}}
\def\gbt{{\tilde {\bar g}}}
\def\ut{{\tilde u}}
\def\vt{{\tilde v}}
\def\vtp{{\tilde v}'}
\def\vtz{{\tilde v_0}}
\def\scrim{${{\cal I}^-} \,\,$}
\def\scrn{{\cal{N}}}
\def\BDV{BDV}
\def\CHR{CHR}
\def\pa{\partial} 
\def\uh{\hat u}

\begin{titlepage}
\bigskip
\bigskip\bigskip\bigskip\bigskip

\centerline{\large \bf Numerical Black Hole Interiors and String
Cosmology Initial Conditions}

\bigskip\bigskip
\bigskip\bigskip

\centerline{\bf R. Madden\thanks{madden@ihes.fr}}
\medskip
{\it Institut des Hautes Etudes Scientifiques, 91440
Bures-sur-Yvette, France} \\ 

\bigskip\bigskip

\begin{abstract}
\baselineskip=16pt
Recent work has proposed the principle of `asymptotic past triviality'
to characterize the initial state in the pre-big bang scenario of string
cosmology, that it is a generic perturbative solution of the low-energy
effective action. Among the more generic sets of solutions which is simple
enough to investigate thoroughly, yet complex enough to exhibit
interesting behavior, is the gravity-dilaton system in spherical symmetry.
Since, in the Einstein frame, this system reduces 
to a massless minimally coupled scalar, 
which has been target of a large body of previous 
investigation, we will draw on this and interpret it in the cosmological
context. Since this scenario necessarily involves the transition from
weak field initial data into the strong field regime, gravitational
collapse, we have made numerical computations to answer some of the 
questions raised on the road to the proposal that `the pre-big bang is 
as generic as gravitational collapse'. 
\vspace{3cm}
\begin{flushleft}
IHES/P/00/71\\
\today
\end{flushleft}
\end{abstract}
\end{titlepage}
\starttext
\baselineskip=18pt
\setcounter{footnote}{0}
\section{Introduction}

Since superstring theory in its various manifestations \cite{Polch}, 
is currently the best candidate for a 
theory incorporating General Relativity (GR) in a quantum context, it
is natural to explore its implications for cosmology, in 
particular what it can say about a proposed early period of
rapid expansion, inflation \cite{inflation}. While string
theory contains GR as a limiting case, the metric tensor is
supplemented by other massless fields, universally including the scalar
dilaton and an anti-symmetric tensor. 
There is additional field content dependent on the
choice of a particular perturbative string theory, including
numerous higher index form fields and moduli \cite{LWC}. 

Of particular importance
is the dilaton, which plays the role of a Brans-Dicke type  
field dynamically controlling the strength of the gravitational coupling. 
While the dilaton
was recognized as a candidate inflaton early on, it was also
realized that it created problems for standard potential
dominated inflation \cite{baddil}. However it was also realized that
a massless dilaton could drive inflationary expansion in the
early universe (mass would be acquired later, 
presumably through a mechanism related to supersymmetry breaking)
by virtue of an instability related solely to its kinetic energy.
Using this as a base, Gabriele Veneziano and collaborators made an 
`heretical' proposal for cosmological initial conditions \cite{PBB},
the pre-big bang (PBB). Since kinetic dilaton-driven inflation naturally pushed
the universe in the direction of increasingly strong coupling and 
increasingly large curvature it was natural to propose that the 
universe began in a regime of weak coupling and small 
curvature, weakly coupled perturbative 
string theory in a nearly Minkowski
vacuum. Furthermore, if a transition to a decelerated evolution and a 
stabilized coupling could be made, this era could be smoothly
joined to a standard radiation dominated cosmology.
While the mechanism of this transition (`graceful exit') remains
obscure, it seems plausible that it could be achieved by
high curvature and strong coupling
corrections to the lowest order effective action \cite{graceful},
which are not presently unambiguously calculable. 

This radically different choice of initial conditions could
not long escape the notice of cosmologists accustomed to thinking
of initial conditions at near Planck scale conditions. In
\cite{TurWein} it was pointed out that the initial spatial
curvature scale of the universe would need to be extremely large
in string units to give sufficient e-folds of inflation, and
\cite{KalLinBou} underlined this objection and quantified it in other
ways, pointing out that it seemed to lead to other unnaturally large or small
numbers and questioning whether the PBB in fact solved the problems 
inflation is supposed to solve.
Part of these objections can be dealt with simply by 
pointing out that the posited PBB initial conditions constitute a 
{\it classical} state, so there is no reason why it should have a 
characteristic scale of the order of the string scale. Indeed the
limit to the 
initial curvature scale is of the same order as its initial expansion
rate \cite{GaspHor}, so there is simply no curvature problem to solve.
Nonetheless, the question of how broad a range of initial conditions
could be regarded as good for PBB inflation remained.
In \cite{BMUV} the authors explored the
idea of a constant negative curvature Milne universe as a universal
past attractor for PBB universes, but as pointed out in \cite{ClaLidTav}
a broadened class of Bianchi types includes solutions exhibiting
gravitational plane waves as past attractors.

This led to the proposal of \cite{BDV}, that a sufficiently generic
state for PBB initial conditions is `asymptotic past triviality' (APT),
a generic perturbative solution to the lowest order effective action.
More concretely, they propose a bath of weak gravitational and dilatonic
radiation. They pointed out that this
initial condition of radiation can still
exhibit the runaway dilaton behavior characteristic
of the PBB in the form of localized gravitational collapse. The interior 
of this collapsing region may serve as a seed (a baby universe).
Just as in the simplest flat PBB, the
runaway inflation produced by the
kinetic energy of the dilaton is hoped, even expected, to be stopped by the
singularity regulating effects expected of strings as extended
objects, a graceful exit. With the abrupt curvature changes would
come copious particle production and a transition to a decelerated
evolution, a hot big-bang, followed by the standard cosmological 
scenario. 

A threat to this general notion of a black hole containing
our universe is
pioneering work by Belinskii, Khalatnikov and Lifshitz (BKL) \cite{BKL}
suggesting that a generic approach to a vacuum singularity in Einstein
gravity exhibits an infinite number of violent anisotropic
oscillations. Encouraging is the observation by two of the same
authors, among others \cite{BK}, that the presence of a scalar
field quenches these oscillations. However, recent work
\cite{DamHen} has suggested that truncating our field content 
to the graviton and
dilaton is unrealistic, and that including the full form field
content of the full string theory will restore the BKL oscillatory
nature of the singularity. 

A modified realization of this idea of realizing PBB initial 
conditions via collapse was proposed in \cite{FKV}. They
suggest a similar scenario in the context of singularity formation
by colliding plane waves. By examining the analytic approach to the singularity
cosmologically in the string frame, they find a significant region
in the parameter space of initial data which supports PBB inflation, in 
the sense of having all directions expanding in the string frame. 
But the effect of including the
antisymmetric tensor field in the plane wave collision
scenario was studied in \cite{VenBoz}, with the conclusion that
this also tends to spoil the inflationary behavior and that PBB 
inflation may not be able to replace altogether more standard forms
of potential driven inflation. Nevertheless, they suggest
it may play a complementary role in providing initial conditions for
subsequent standard inflation, so singularity onset from
perturbative initial conditions may still play an important
part in the formation of the universe.

Restricting the field content to the graviton and the dilaton and
applying a conformal rescaling from the string frame leads to a
physically equivalent picture in the Einstein frame, gravity coupled
minimally to a massless scalar. Further restriction to spherical
symmetry gives a framework which is both theoretically and numerically
tractable, but is still thought to retain the essence of physically
realistic gravitational collapse. A long series of works by
Christodoulou used this system to prove rigorous results ranging
from the existence of global solutions to Einstein's equations, to
criteria guaranteeing the collapse of initial data to his most
recent investigations of the genericity of naked singularities
\cite{Chr1}-\cite{Chr6}. Further evidence of the richness of the system comes
from the discovery of universal scaling and various sorts of critical
phenomena \cite{Cho}.

We will concentrate on some questions raised in \cite{BDV}. 
To probe the system the authors develop weak and strong field expansions for
the evolution. However matching them only seems possible
numerically, and we have developed an algorithm to do
this matching. Our main goal is to explore the degree to which
the black hole interior can resemble the dilaton driven regime of the
PBB scenario. Other numerical studies of the PBB have focused on
questions of its perturbative stability \cite{numpbb}, but we
will study the full nonlinear evolution.
We will also look at a collapse criterion proposed
in \cite{BDV}
on the basis of a weak field expansion and attempt to draw 
some comparison with a more rigorous bound of Christodoulou.

\section{The Action, Conformal Frames and EOMs}
\label{genevolve}
\setcounter{equation}{0}

The tree-level low-energy effective action of a generic string
theory, truncating to the graviton-dilaton content is 
\beq
S_s = \frac{1}{\alpha^\prime} \int d^4 x \, \sqrt{G} \, 
 e^{-\Phi} [R(G) + G^{\mu \nu} \partial_{\mu} 
\Phi \, \partial_{\nu} \Phi] \,,
\label{saction}
\eeq
where $G$ is the string frame metric, $\Phi$ is the four dimensional
dilaton and $\alpha'$ is related to the fundamental string
length scale by $\alpha' \sim {\ell_s}^2$. 
While this is the
physical frame in string theory and it is in this frame that the
inflationary behavior of the dilaton is manifested most clearly,
we will find it convenient to work in the conformally related but physically
equivalent Einstein frame. Defining
\beq
g_{\mu \nu}=e^{-\Phi} G_{\mu \nu},
\label{strtoein}
\eeq
this action becomes
\beq
S_e = \frac{1}{\alpha^\prime} \int d^4 x \, \sqrt{g} \, 
 [R(g) - {1 \over 2} g^{\mu \nu} \partial_{\mu} \Phi 
\, \partial_{\nu} \Phi] 
\,,
\label{eeaction}
\eeq
the usual action for gravity coupled to a massless scalar.

Reducing to spherical symmetry gives a numerically tractable but
still rich system.
We use an algorithm based on a form of the scalar-Einstein equations
in radial symmetry given in a work of Christodoulou \cite{Chr1}.
Its use as the basis for a numerical code was pioneered in
\cite{GolPir} and has since been exploited by many authors
\cite{Gar}. It leads naturally to a completely characteristic 
stepping code. As such it has the same capabilities as double null codes
\cite{BurOri,HamSte}, although it is most naturally expressed in a
mixed coordinate system $(u,r)$, where $u$ is a null coordinate and $r$ is 
the area $r$ coordinate (spheres at fixed $r$ have area $4 \pi r^2$). Though
thought of as a coordinate in the definition of the metric, in the
formulation of the algorithm $r$ will 
become a quantity to be evolved along a second null direction. We 
deliberately do not specify whether $u$ corresponds to an ingoing or outgoing
null direction, since will have occasion to use it both ways. Use of this
algorithm along ingoing nulls was made in \cite{BraSmi} to explore
charged singularity structure.

Through the introduction of an auxiliary variable we reduce the second
order field equations to a first order system. Further, several of the
field equations will be enforced by quadratures along nulls. Thus the
algorithm requires no numerical differentiation during the
evolution. This leads to an extremely stable `pocket-sized' code,
accounting for its popularity. Finally the auxiliary function will
turn out to be proportional to the `news' function at past null
infinity, facilitating comparison with \cite{BDV}.

We express the metric in the $(u,r)$ coordinate system as
\beq
ds^2=-g(u,r) \gb(u,r)~ du^2-2 g(u,r)~ du~ dr+r^2 d\omega^2.
\label{nullmet}
\eeq
In terms of the scalar normalization of \cite{Chr1}, 
the Einstein-scalar equations are
\beq
R_{\mu \nu}=8 \pi \nabla_{\mu} \phi(u,r) \nabla_{\nu} \phi(u,r),
\label{eom1}
\eeq
together with the scalar field equation
\beq
\nabla^{\mu} \nabla_{\mu} \phi=0,
\label{eom2}
\eeq
where $\phi$ is related to $\Phi$ by
\beq
\Phi=4 \sqrt{\pi} \phi.
\eeq
For notational reasons we will also write $\hb=\phi$. We then define
the auxiliary variable
\beq
(r \hb)_{,r}=h. \label{hdef}
\eeq
To connect this with the notation of \cite{BDV}, we recall that
they define the `news' function in the following way. The free 
field behavior of initial perturbative data is given by
\beq
\Phi={{f(v)-f(u)} \over {r}}, \label{weakfield}
\eeq
with $u=t-r$ and $v=t+r$. The news is then defined by
\beq
\scrn(v)=f'(v).
\label{news}
\eeq
Putting this together and correcting for normalization, we see that
\beq
2 \sqrt{\pi} h \rightarrow \scrn
\eeq
as $u \rightarrow -\infty$ at constant $v$ (past null infinity or \scrim).

Finally we define the differential operator
\beq
D_u(f)=f_{,u}-{\gb \over 2} f_{,r} \label{duop}.
\eeq
From the form of the metric (\ref{nullmet}), this is differentiation
in the null direction complementary to the nulls of constant $u$.
That is,
\beq
D_u(f)=\left (\frac{\partial f}{\partial u}\right )_v \label{duop2},
\eeq
where $v$ is a complementary null coordinate to $u$. 
The component content of the equations of motion (\ref{eom1}) 
and (\ref{eom2}) can now be written very simply:
\beq
(\log g)_{,r}&=& 4 \pi {(h-\hb)^2 \over r} \label{greq} \\
(r \gb)_{,r}&=&g \label{gbreq} \\
D_u(h)&=&{(g-\gb)(h-\hb) \over 2 r} \label{hev} \\
0={\cal{L}}&=&{(g-\gb) \over 2 r}+D_u \left[\log({\gb \over g})\right]
 -{8 \pi r (D_u \hb)^2 \over \gb} \label{cons}. 
\eeq
As promised all derivatives are taken along null directions, so 
we now begin to think of all functions as functions of $(u,v)$,
though we will abuse notation somewhat and sometimes
write them as functions of $(u,r)$, which presents no problem
since $r(u,v)$ will be monotone in $v$.
As usual, conservation imposes a redundancy on the equations of 
motion and one can be regarded as superfluous if we insure it is
satisfied `initially'. We choose to regard the most complicated
equation (\ref{cons}) as the constraint and use the others to 
construct an algorithm. 
Some tedious work of differentiating and plugging
in the other eoms shows,
\beq
{\cal{L}}_{,r}={-2 {\cal{L}} \over r} \label{drcons}.
\eeq
This has the solution ${\cal{L}}=C/r^2$, implying 
that if the constraint is satisfied at any single point along the null,
it will be satisfied everywhere along the null 
provided the other equations of motion are
satisfied. 

We choose an initial $u=u_0$ null. We specify initial data for
$r$ and $h$ on a grid of points in $v$ along the null. The choice 
of the $v$ coordinate is completely arbitrary. From this data
(\ref{hdef}) is used to construct $\hb$. We can then use 
(\ref{greq}) to construct $g$ and (\ref{gbreq}) to construct
$\gb$. Clearly initial conditions will be required somewhere 
along the null, but once this is done the constructions can 
carried out by simple quadrature. The choice of these initial 
conditions will be dictated by the need to solve the constraint.
Once we have $r$, $h$, $\hb$, 
$g$, and $\gb$ we can use (\ref{hev}) to step $h$ to the next
null at $u_0+\Delta u$. We use the obvious evolution operator
for $r$ following from (\ref{duop}),
\beq
D_u(r)={-\gb \over 2} \label{rev},
\eeq
to step the $r$ values on the grid to the next null. Having
new values of $r$ and $h$ on the next grid allows us to iterate the
process.

The location of the apparent horizon is given by the solution
to the equation
\beq
\nabla_{\mu} r \nabla^{\mu} r={\gb \over g}=0.
\eeq
Upon crossing this surface even the outgoing nulls become `ingoing',
in the sense of having a decreasing value of $r$ along the null,
signalling the formation of a trapped surface and inevitable 
collapse. Finally the Bondi mass function is given by
\beq
m={r \over 2} (1-{\gb \over g}),
\eeq
so the apparent horizon can also be characterized by the condition
$2 m/r=1$.

\section{Stepping Modes}
\setcounter{equation}{0}

This formulation of characteristic evolution, while technically simple,
cannot cover the whole of the collapse spacetime in a single
coordinate patch for several reasons. First, the metric components
being evolved, $g$ and $\gb$, are specific to an ingoing or outgoing 
null coordinate system. In an outgoing null coordinate system 
these components will diverge on approach to the
apparent horizon, signalling a breakdown
in the coordinate system. A global ingoing null system does not 
have this problem, but is at best awkward to attach to $r=0$, even
before the formation of the $r=0$ singularity, due to mixed
boundary conditions. Finally,
on approach to \scrim, $r$ will diverge.
The primary problem we wish to examine, the relationship of asymptotic
news to asymptotic behavior near the singularity involves 
approaching all of these limits.

To deal with this we split the spacetime into three regions and
develop variants of the basic algorithm described in section
\ref{genevolve}. The first ($u$-mode) is one based on outgoing nulls
and imposing boundary conditions at the regular $r=0$ origin. The
second ($v$-mode) is based on ingoing nulls and takes boundary
conditions on an outgoing null boundary from data extracted from the
$u$-mode variant, allowing us to approach the $r=0$ singularity. 
Finally in the third (\scrim-mode), we rescale the equations
to put the \scrim boundary at a finite position on the grid. Next
we describe these modes in more detail.

\subsection{$u$-mode Stepping on Outgoing Nulls Touching $r=0$}
\label{uevolve}
\setcounter{equation}{0}

\begin{figure}[t]
\centerline{\epsfig{file=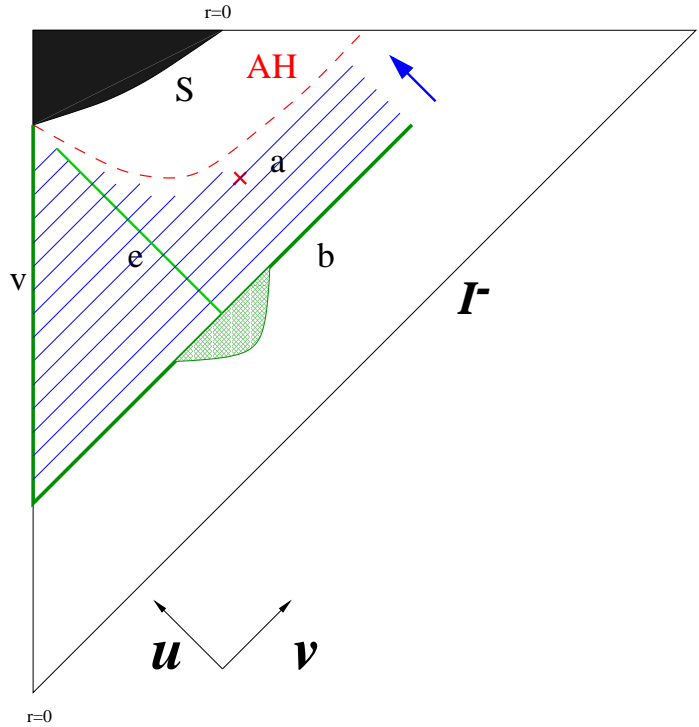,width=0.6\textwidth,angle=0}}
\caption{\em\baselineskip=11pt Evolution along outgoing nulls in a 
portion of a collapse spacetime. Shown are the singularity {\rm S},
the apparent horizon {\rm AH} and \scrim. Initial conditions are
fixed by regularity at the central axis {\rm v} ($r=0$) 
and initial data $r$ and $h$
are supplied
along the outgoing null {\rm b}. The arrow indicates the direction of
null stepping. Beginning at point {\rm a} the stepped ingoing nulls
are truncated near the apparent horizon. Initial data is then
extracted along null {\rm e} for use in the ingoing
null version of the algorithm.}
\label{figu}
\end{figure}

Here we include $r=0$ as the initial point in the grid and identify
$u$ as a null coordinate which is constant along outgoing nulls,
Fig.~\ref{figu}. We will fix our initial data by specifying $h(u_0,v)$
and $r(u_0,v)$ along an initial outgoing null at finite $u=u_0$ 
(`b' in Fig.~\ref{figu}). Regularity at $r=0$ (`v' in
Fig.~\ref{figu}) requires that $h(u,r=0)=\hb(u,r=0)$ (\ref{hdef}) and
$g(u,r=0)=\gb(u,r=0)$ (\ref{gbreq}).  Further we will gauge fix $u$ to
be proper time at the origin $r=0$, so $g(u,r=0)=\gb(u,r=0)=1$. In
light of (\ref{drcons}), we see that regularity at the origin implies
${\cal{L}}=0$, so the  constraint is satisfied everywhere as observed
in \cite{Chr1}.  We note the explicit quadratures of (\ref{hdef}),
(\ref{greq}) and (\ref{gbreq})  subject to these boundary conditions,
\beq
\hb(u,v)&=&{1 \over r(u,v)} {\int_0}^{r(u,v)} h(u,r') dr' \label{phb} \\
g(u,v)&=&\exp \left[ 4 \pi {\int_0}^{r(u,v)} {(h(u,r')-\hb(u,r'))^2
\over  r'} dr' \right] \\
\gb(u,v)&=&{1 \over r(u,v)} {\int_0}^{r(u,v)} g(u,r') dr'. \label{expquad}
\eeq

The choice of $v$ coordinate along the initial null is arbitrary and
serves only to identify which grid point belongs to  which ingoing
null. In practice $v$ is generally taken from a \scrim evolution (see
section \ref{scrimevolve}) so it corresponds to the usual $v \sim t+r$
at past null infinity. The evolution of $r$ and $h$ to the
next null is now carried out using (\ref{hev}) and (\ref{rev}).

As we evolve, the $r$ value of grid locations will decrease eventually
becoming negative. The point then becomes unphysical and is dropped
from the grid and a new $r=0$ point is interpolated. This is a 
consistent procedure since the quadratures (\ref{expquad}) will
maintain the correct regularity boundary conditions at the
new origin.
It is also reassuring to consider the limiting case of the 
evolution of extremely weak news data, the flat space wave
equation. In this limit $u=t-r$, $v=t+r$ and 
$g=\gb=1$ so $D_u(h)=0$. Thus $h(u,v) \propto \scrn(v)=f'(v)$. 
Substituting this into (\ref{phb}) and correcting for normalizations gives
\beq
\Phi(u,v)={2 \over r} {\int_0}^r \scrn(v(u,r)) dr={1 \over {r}} {\int_u}^v
f'(v') dv'={{f(v)-f(u)} \over {r}},
\eeq
so we correctly reproduce the weak field evolution (\ref{weakfield}).
In terms of our stepping algorithm this corresponds to our procedure
of simply discarding points as they pass to negative values of $r$.

We also monitor the value of
\beq
\mu={2 m \over r}=1-{\gb \over g}
\eeq
along the grid.  If $\mu$ approaches 1 (the point `a'
in Fig.~\ref{figu}), then we are approaching the apparent
horizon. This version of the algorithm cannot proceed past this point
since $r$ begins to decrease along
an outgoing null as it crosses the apparent horizon, so the $(u,r)$
coordinate system is not even one-to-one. The value of $r$ at the
grid point where this limit was exceeded is noted as an estimate of
the final black hole size $r_{BH}$, as is an estimate of 
the asymptotic Bondi mass,
the value of $m$ in the  outermost grid point.

We now proceed with the evolution by simply discarding the outer parts
of the grid where a limit on $\mu$ is exceeded (notice $\mu=0$ at
the regular $r=0$ origin as can be seen from a power series expansion
of the functions in $r$).  
This allows us to approach close to the appearance of the
singularity at $r=0$. The integration is halted when the outermost
value of $r$ is less than some $r_{min}$ and we extract initial
data ($u$, $r$ and $\hb$) along the ingoing null containing
that point (`e' in Fig.~\ref{figu}) for use as initial conditions
for  the ingoing null phase of the evolution (see section
\ref{vevolve}).

\subsection{$v$-mode Stepping on Ingoing Nulls Excluding $r=0$}
\label{vevolve}
\setcounter{equation}{0}

\begin{figure}[t]
\centerline{\epsfig{file=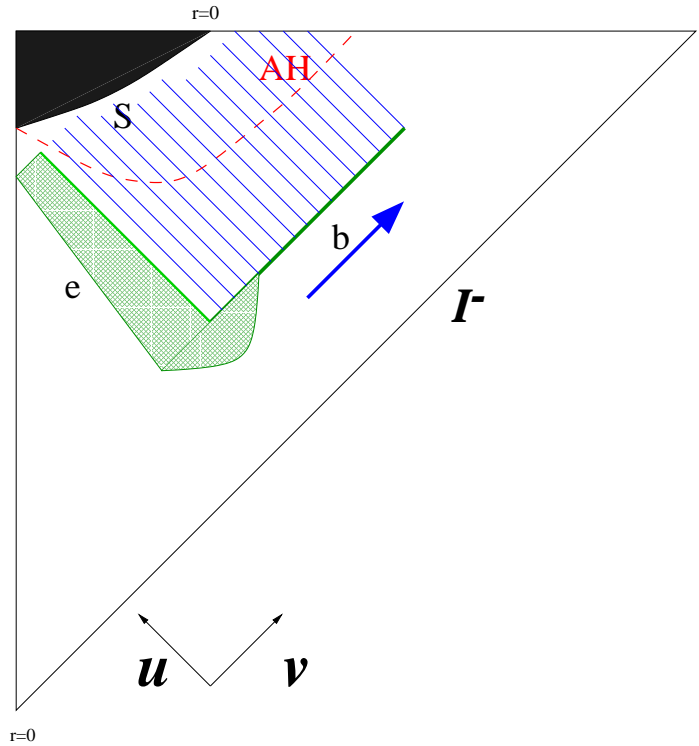,width=0.6\textwidth,angle=0}}
\caption{\em\baselineskip=11pt Evolution along ingoing nulls in a 
collapse spacetime as in Fig.~\ref{figu}. 
Initial data from the $u$-mode evolution is
supplied along the ingoing null {\rm e} and the outgoing null
{\rm b}. The arrow indicates the direction of null evolution.
The ingoing
nulls are truncated as they approach the $r=0$ singularity {\rm S}.}
\label{figv}
\end{figure}

In this mode we abandon our dependence on regularity at $r=0$. Indeed,
generally $r=0$ will be a singular point. The basic algorithm is
identical with that described in section \ref{genevolve} but we now 
reinterpret the `$u$' coordinate 
to be a conventional $v$ coordinate constant along
ingoing nulls. 
Our region of integration is now bounded by an ingoing
null and an outgoing null, Fig.~\ref{figv}. We use the extracted data
from the previous $u$-mode integration to provide initial data on the
ingoing null (`e' in Fig.~\ref{figv}), $u$, $r$ and  $h$ (which is
obtained from the extracted $\hb$ by numerical  differentiation).
On the outgoing null
boundary (`b' in Fig.~\ref{figv}) we similarly extract values for $r$,
$\hb$, $g$ and $\gb$, but change $g \rightarrow -g$ and
$\gb \rightarrow -\gb$ corresponding to the gauge change $v
\rightarrow -v$. This gives us a $v$ coordinate increasing along the
outgoing null as implicitly defined by the analog of (\ref{rev}),
\beq
D_v(r)={-\gb \over 2} \label{rev2}.
\eeq
Equations (\ref{hdef}), (\ref{greq}) and (\ref{gbreq}) are solved using the
initial data on the outgoing null, instead of using the regularity
conditions at $r=0$.

We now evolve just as before, but in the $v$ direction, stepping the
data on the ingoing initial null outwards, employing the same
equations as in section \ref{genevolve}
with $v$ substituted for $u$. These ensure
the solution of all of Einstein's equations except for the constraint
(\ref{cons}). But as we have argued through (\ref{drcons}), we only
need to show the constraint is satisfied at one point along each
ingoing null. We choose this point to be its intersection with the
outgoing null boundary. So we wish to show
\beq
{(g-\gb) \over 2 r}+{D_v(\gb) \over \gb}-{D_v(g) \over g} -{8 \pi r
(D_v \hb)^2 \over \gb}=0 \label{vcons}
\eeq
along the outgoing null, where
\beq
D_v(f)=\left (\frac{\partial f}{\partial v}\right )_u.
\eeq
By virtue of our definition of $v$ coordinate (\ref{rev2}) we can
replace the $D_v$ operator in (\ref{vcons}) by
\beq
D_v(f)={-\gb \over 2} D_r(f)
\eeq
where 
\beq
D_r(f)=\left (\frac{\partial f}{\partial r}\right )_u.
\eeq
So $D_r(f)$ is the same as $f_{,r}$ in the $(u,r)$ coordinate
system.
Now from (\ref{hdef}), (\ref{greq}) and (\ref{gbreq}) 
respectively (which were used to construct our original outgoing null
data), we have,
\beq
D_r(\hb)&=&{\hat h-\hb \over r} \\
{D_r(g) \over g}&=& 4 \pi {(\hat h-\hb)^2 \over r} \\
D_r(\gb)&=&{g-\gb \over r}.
\eeq
We have written $\hat h$ instead of $h$ to emphasize that it refers
to the function $h$ in the old coordinates $(u,r)$. 
$h$ and $\hat h$ are related
to the invariant $\hb=\phi$ by $r$ derivatives along different null
directions. Substitution of these results into (\ref{vcons}) yields
an identity. So we conclude we have consistent data for an integration
of the eoms. 

According to a theorem of Christodoulou \cite{Chr3}, the apparent 
horizon will extend outwards from the regular $r=0$ axis 
in an `achronal'
manner, spacelike except for possible null parts. So provided we
have penetrated deep enough in $r$ with our ingoing initial null the
evolving integration region should encounter the apparent horizon
and pass inside.
Notice that, unlike in the $u$-mode scheme, $\gb$ can pass through zero
and change sign. $\gb=0$ signals the apparent horizon and there is 
no divergence in the metric functions in the $v$-mode gauge. Once an
outgoing null crosses the horizon its $r$ value begins to decrease as
it falls in towards the singularity and we truncate the grid again at 
some fixed $r_{min}$.  

Once the horizon appears we have grid points inside falling towards the
singularity and those outside flying away, so the region around the 
horizon quickly empties. We replenish them using
a `point-splitting' scheme \cite{BurOri}, simply interpolating new points
to keep the grid populated. Also the $u$ coordinate grid
becomes extremely flat. To integrate to large values of $v$ we would
need a `gauge-correction' and to increase efficiency a `point-removal'
scheme to eliminate unnecessary points. 
The extent of the region of spacetime we are 
interested in has not required these steps.

\subsection{\scrim-mode Outgoing Null Evolution Excluding $r=0$}
\label{scrimevolve}
\setcounter{equation}{0}

\begin{figure}[t]
\centerline{\epsfig{file=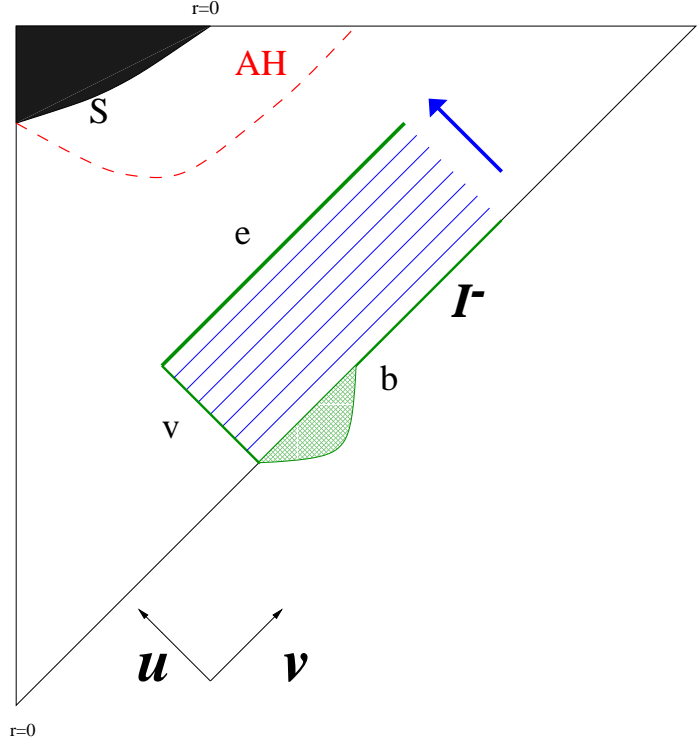,width=0.6\textwidth,angle=0}}
\caption{\em\baselineskip=11pt Evolution along outgoing nulls from
\scrim in a collapse spacetime as in Fig.~\ref{figu}. 
Initial data is given along
a portion of \scrim, {\rm b}, and vacuum initial data is set 
along the ingoing null {\rm v}. The arrow indicates the 
direction of null stepping. Data at finite $r$ is 
extracted along a final outgoing null {\rm e}.}
\label{figi}
\end{figure}

Finally, we introduce one more operating mode to bring past null infinity, 
\scrim, into the grid, Fig.~\ref{figi}. 
We will rescale the auxiliary functions involved in 
the $u$-mode integration in such
a way that derivatives and functions are well defined and finite at \scrim.
To this end we introduce a compactified $u$ coordinate and $v$ related 
quantity,
\beq
\ut&=&{-1 \over u} \\
\vt&=&u+2 r=2 r-{1 \over \ut}.
\label{vgauge}
\eeq
$v$ will be gauge fixed to equal to $\vt$ at \scrim ($\ut=0$). 
This is, as usual,
an arbitrary choice since it plays no essential role other than being a label,
but (\ref{vgauge}) ensures that $v$ approaches the usual sort of
$v \sim t+r$ coordinate at \scrim.
We also restrict ourself to initial data such that $\scrn(v)=0$ for
$v<v_0$.
This allows us 
to impose vacuum boundary conditions along the ingoing boundary null
$v=v_0$ (`v' in Fig.~\ref{figi}) so
we are not obliged to include the origin on the grid. To be clear we 
work in the $(\ut,v)$ null coordinates, though again we occasionally
abuse the notation and regard functions as depending on $(\ut,\vt)$.
$\vt(\ut,v)$ will play a role 
similar to $r$ in $u$-mode, it will be evolved. Initial data
specified at \scrim are $\vt(0,v)$ and $\htt(0,v)$. We now define the
auxiliary functions, which are rescaled versions of (\ref{expquad})
modified to contain an inner cutoff radius,
\beq
\htt(\ut,v)&=&h(u,v) \\
\hbt(\ut,v)&=&u \hb={-1 \over (1+\ut \vt)} \int_{\vt(\ut,v_0)}^{\vt(\ut,v)} 
  \htt(\ut,\vtp) d\vtp \\
\gt(\ut,v)&=&u (g-1)={-1 \over \ut} \left[ \exp{(4 \pi \ut 
\int_{\vt(\ut,v_0)}^{\vt(\ut,v)} 
  {(\htt(\ut,\vtp)+\ut \hbt(\ut,\vtp))^2 \over (1+\vtp \ut)} d\vtp)}-1 \right] 
\label{gteqn} \\
\gbt(\ut,v)&=&u^2 (\gb-1)={-1 \over (1+\ut \vt)}  
\int_{\vt(\ut,v_0)}^{\vt(\ut,v)}
  \gt(\ut,\vtp) d\vtp.
\eeq

In terms of these variables the evolution equations read
\beq
D_\ut&=&u^2 D_u \\
D_\ut \vt&=&-\gbt \\
D_\ut \htt&=&{{-(\gt+\ut \gbt)(\htt+\ut \hbt)} \over (1+\ut \vt)}.
\eeq
\scrim ($\ut=0$) is now a regular null on the grid. Explicitly,  
at \scrim (\ref{gteqn}) can be replaced by the only non-vanishing term in its
power series expansion in $\ut$,
\beq
\gt(0,v)&=-4 \pi \int_{\vtz(0,v_0)}^{\vt(0,v)} \htt(0,\vtp)^2 d\vtp. 
\eeq

The evolution is halted when the innermost point of the data reaches
a specified finite $r$ and data is extracted. It can then
be connected through
vacuum to the origin and evolution is continued in $u$-mode.

\section{Numerics, Tests and Verifications}
\setcounter{equation}{0}

We use the lowest order algorithms still compatible with quadratic
convergence. The $r$ integrations along the nulls are carried out
using the trapezoid rule, the null stepping is carried using a 
second order Runge-Kutta method and linear interpolation is used
where necessary. Besides the advantage of simplicity, these methods
work tolerably well even with data of 
low differentiability (e.g. \cite{BDV} (4.37)). 

It is useful to choose an analytic test case exhibiting features
in common with scalar collapse. A simple case is spatially flat
FRW scalar collapse, the simplest version of the PBB. 
While too simple to provide a good test
of an $(r,t)$ code, having no spatial gradients, when written in 
radial null coordinates we will see it has features in common with
radial scalar collapse and has non-zero gradients in the null
directions.

We write the FRW metric
\beq
ds^2=-dt^2+a(t)^2 (dR^2+R^2 d\omega^2)
\eeq
in terms of conformal time $\eta$,
\beq
ds^2&=&a(\eta)^2 (-d\eta^2+dR^2+R^2 d\omega^2) \\
\eta&=&\int^t {1 \over a(t')} dt'.
\eeq
Introducing null coordinates defined by conformal time at
the origin, $\uh=\eta-R$ and
$v=\eta+R$, we can write the PBB solution,
\beq
a(\eta)&=&\sqrt{\eta \over \eta_0},\,\,\eta<0 \\
\phi(\eta)&=&{-\sqrt{3} \over {2 \sqrt{\pi}}} \log(a(\eta)),
\label{exactpbb}
\eeq
in terms of these null coordinates. Finally we replace the $\uh$
coordinate with the $u$ coordinate defined in terms of proper 
time at the origin
\beq
\uh=\eta_0 \left[ {u \over u_0} \right]^{2 \over 3},
\eeq
where $\eta_0=3 u_0/2$ is a chosen initial time ($a(\eta_0)=1$). 
Finally we change to the area $r$ coordinate by $R={r/a(u,v)}$.

The resulting solution has divergent news at \scrim, requiring us to 
specify initial data at finite $u$, but the rest of the algorithm
proceeds just as with a realistic collapse since this solution has
a spacelike singularity at $v=-\uh$ ($\eta=0$) 
hidden by a spacelike apparent 
horizon at $v=-\uh/3$, allowing us to test the horizon crossing
and $v$-mode approach to the $r=0$ singularity.

\begin{figure}[t]
\centerline{\epsfig{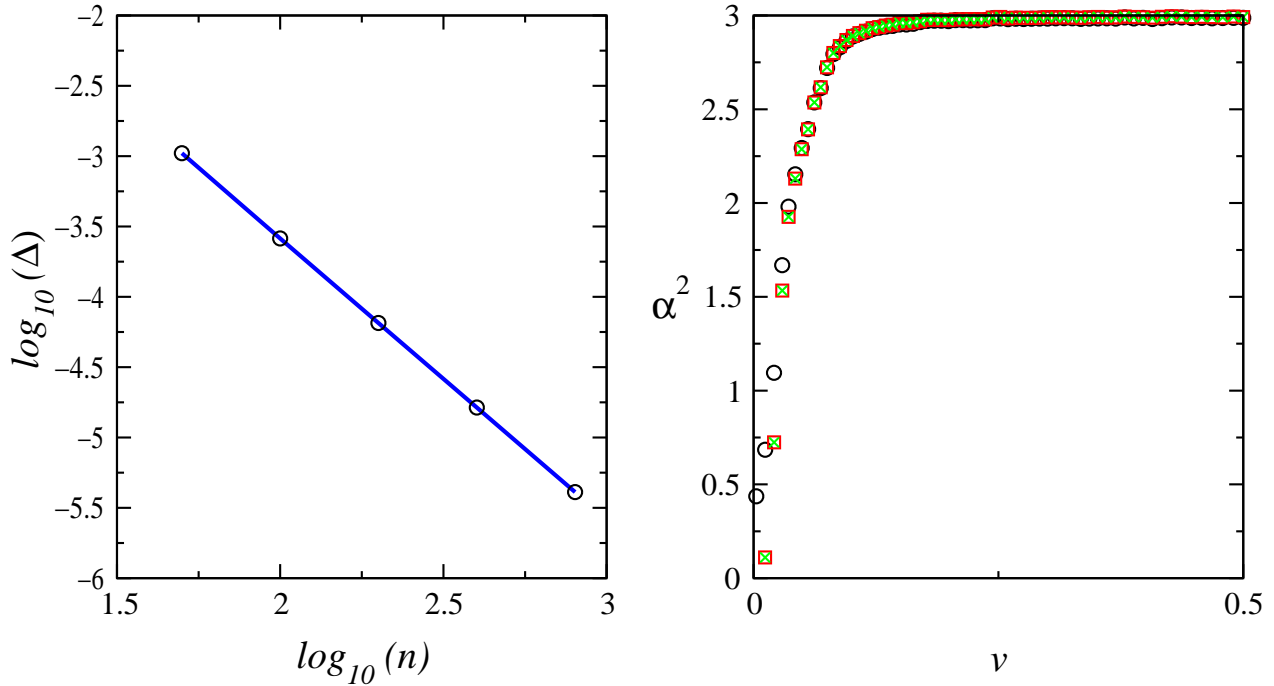}}
\caption{\em\baselineskip=11pt 
Left: the log of the maximum difference, $(\Delta)$, 
of the numerical integration of $\phi$ from the exact PBB solution 
(\ref{exactpbb})
plotted versus the log of the number of grid points $n$ for 
$n=50,100,200,400,800$. Right:
a numerical extraction of the asymptotic value of $\alpha^2(v)$ for 
the same solution from
$\phi$ ($\circ$), $m$ ($\Box$) and $g$ ($\times$) over a range of $v$.}
\label{figconv}
\end{figure}

First we verify the order of convergence of the algorithm by
checking it against the exact solution (\ref{exactpbb}). We 
take initial $r$ and $h$ data from the exact solution 
at $u_0=-2$, $-2<v<0.5$.
The data is then evolved to $u=-1$ and the results are compared 
with the exact solution. Fig.~\ref{figconv} shows the logarithm
(base 10) of the absolute
value of the largest error, $\Delta$, in $\phi$ on the 
grid versus the 
logarithm of the number of initial grid points $n=50,100,200,400,800$. 
It is clearly
almost exactly quadratically convergent. 

The full algorithm is then deployed to test the near $r=0$ approach
to the singularity. Anticipating the asymptotic analysis of 
section \ref{secasym}, we numerically extract the small $r$ behavior of
the dilaton near the singularity. As we will discuss this is 
expected to be of the form $\Phi(v)=2 \alpha(v) \log(r)$, 
and since this is homogeneous and isotropic collapse, the value of
$\alpha(v)=-\sqrt{3}$ is expected. Using 400 points on the 
grid, the horizon first appears 
at $r=0.42$ (marked by $\mu>0.98$) 
and the $v$-mode integration is pushed to $r=0.001$.
The results are shown in the second figure in
Fig.~\ref{figconv}. We show the values of $\alpha^2(v)$ extracted from
the asymptotics of $\Phi$, $m$ and $g$. The small $v$ deviation from
the asymptotic values are expected, since the null geodesics have
not yet penetrated close to the singularity, see Fig.~(\ref{figv}). 
In this particular run we 
find $\alpha(v) \rightarrow -1.728$ compared with $-\sqrt{3}=-1.732$,
confirming we can reach near enough to the singularity to extract
asymptotic values of $\alpha(v)$. As we will discuss further in
section \ref{secasym}, this is the quantity which will characterize
the cosmological nature of the collapse.

Among other tests of the code, 
we repeat a calculation of the collapse of 
a initial pulse in $\phi$ and compare with the data presented in \cite{Bur} 
with good agreement. We note that although his integration
proceeds several orders of magnitude deeper into the black hole 
than we do, he finds almost no change in the asymptotic $\alpha(v)$
over the last few orders of magnitude, giving us some confidence we
can get a good idea of the asymptotic behavior with a relatively 
shallow penetration inside the apparent horizon. 
We also compare with other numerical results (e.g. \cite{GolPir}) 
and other exact solutions. In addition we verify agreement between
the various stepping modes on regions of overlap and numerically 
evaluate the constraint. All tests indicate
the code produces stable integrations in agreement with others 
in the literature.

\section{Collapse Criteria}
\setcounter{equation}{0}

In attacking the question of the structure of scalar collapse,
the authors of \cite{BDV} develop a perturbative expansion of 
the weak-field behavior in terms of powers of the strength of 
the news function, as defined by (\ref{news}).
They obtain
a concise result for the quadratic approximation of
the mass ratio $\mu=2 m/r$ for news data $\scrn(v)=f'(v)$,
\beq
\mu_2 (u,v) = \frac{1}{v-u} \int_u^v dx \, f'^2 (x) - \left [ 
\frac{f(v) - f(u)}{v-u} \right ]^2 \, .
\label{bdvcrit}
\eeq
They further simplify by noting that if one defines
the average of a function by
\beq
\langle g \rangle(u,v)={1 \over v-u} \int_u^v g(x) dx,
\eeq
and defines a variance by
\beq
Var(g)(u,v)=\langle (g-\langle g \rangle (u,v))^2 \rangle (u,v),
\eeq
this becomes just
\beq
\mu_2(u,v)=Var(\scrn)(u,v).
\eeq

Since $\mu=1$ is the signal of the formation of an apparent horizon
leading to inevitable collapse,
this leads them to define the maximum of its quadratic approximation
\beq
\BDV(\scrn)=\sup_{u,v \atop u \leq v} Var(\scrn)(u,v),
\eeq
and state that news data will collapse if it satisfies an 
inequality of the form
\beq
\BDV(\scrn) \geq C(\scrn),
\eeq 
where $C(\scrn)$ depends only on the shape of the news and
not on its amplitude. So $C(\scrn)$ measures one aspect of 
the departure from linearity of the evolution of $\scrn$.
The usefulness of a criterion of this form is measured by the 
complexity of the functional $C(\scrn)$ and would be maximized
if it could be replaced by a constant. But it could still be
of some use if $C(\scrn)$ would be bounded for a family of 
`reasonably generic' initial data and perhaps only depend weakly on
$\scrn$. 
Aside from its simplicity the criterion
is suggestive because it shares the exact scale invariance of 
the massless gravity-scalar system ($r \rightarrow k r$).
The perturbative expansion leads them
to estimate $C(\scrn)=O(1)$. They also examine an exact collapse,
\cite{roberts}, and find $C(\scrn)=1/4$ for this particular case.

To examine this numerically, we select a family of news data depending
on a number of parameters, map the boundary between collapse and
dispersal in this parameter space and then evaluate 
$\BDV(\scrn)$ along this boundary.
Specifically we define a `wavelet',
\beq
n(v)=a \cos({{2 \pi (v-v_0) f} \over w}) \exp(-{(v-v_0)^2 \over w^2}).
\label{wavelet}
\eeq
Assigning $\scrn(v)=n(v)$ we see that this initial data has four
parameters, amplitude $a$, frequency $f$, width $w$ and center $v_0$. 
$v_0$ is clearly irrelevant since 
translation of news data just shifts the value of $u$ where the
pulse approaches the origin. The irrelevance of $w$ is less 
obvious, but comes from the scale invariance of the system. 
So shifting $w$ can change the
size of black hole formed, but cannot affect whether it will form.
We have verified the numerics show this same insensitivity. So we will 
set these variables arbitrarily ($v_0=-2$, $w=1$). 
We then choose a value for $f$ and locate
a value of $a$ where the data disperses and a larger value of $a$ where
it collapses. Searching by bisection we find the critical
collapse amplitude $a_{0}(f)$ for this value of $f$. Doing this for
a number of values of $f$ allows us to collect a set of critical values
of $a_{0}(f)$ for different values of $f$. 
Finally calculating $\BDV(\scrn)$ for these critical collapse news
functions gives us a cross-section of the shape of $C(\scrn)$ across
this family of functions.

\begin{figure}[t]
\centerline{\epsfig{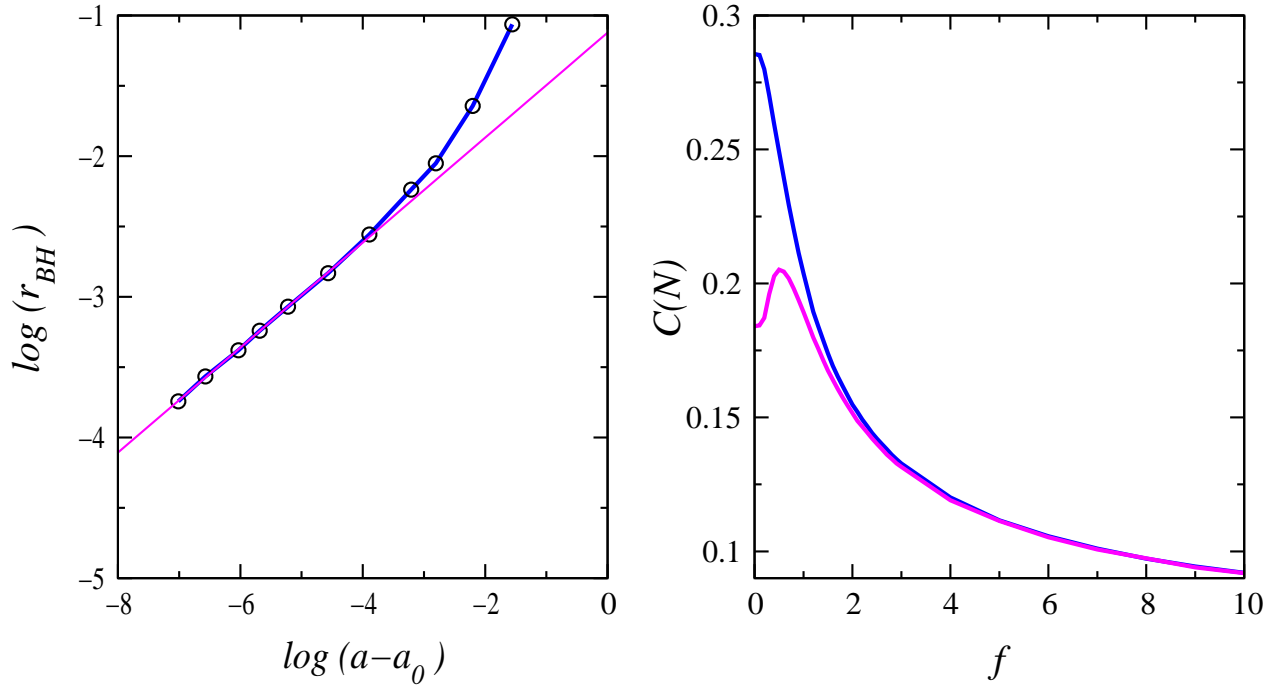}}
\caption{\em\baselineskip=11pt 
Left: Choptuik scaling for $f=0$, $w=1$ news ($\scrn(v)=n'(v)$ 
(\ref{wavelet})). Log of the black 
hole radius versus log of the difference between the amplitude $a$ 
and the critical amplitude $a_0=0.1896$. A best fit line to the
first five points is also shown with slope 0.373. Right: The $C(\scrn)$
functional appearing in the $\BDV$ criterion as a 
function of frequency $f$. On the upper trace $\scrn(v) \propto n(v)$
and on the lower $\scrn(v) \propto n'(v)$.}
\label{crit}
\end{figure}

Before doing this we check that the algorithm can accurately approach the 
collapse threshold by extracting
the Choptuik exponent of the scaling of the black hole radius with
the difference in amplitude from the critical amplitude 
\beq
r_{BH} \propto (a-a_0)^\gamma,
\eeq
where the constant $\gamma$ is expected to depend only on the 
type of collapsing matter and not on the details of the 
waveform \cite{Cho}.
The results for $f=0$ news data are shown
are shown in the first graph in Fig.~\ref{crit}. A best fit line
to the first five points yields a slope of $\gamma=0.373$. 
This can be compared
with other values in the literature, 0.376 in the first
reference in \cite{Cho} or 0.374 in \cite{HamSte}.

We will use the wavelet data $n(v)$ as initial data $\scrn(v)$.
We also take a second family of news functions $\scrn(v) \propto n'(v)$ 
While looking substantially 
different for low frequencies where the derivative of $n(v)$ is dominated
by the gaussian component, at higher frequencies 
these functions will differ mainly 
by a phase coming from the oscillatory component. The dependence
of $C(\scrn)$ as a function of $f$ for these two families is shown
in the second graph in Fig.~\ref{crit}. 

As one might expect, the phase difference at higher frequencies makes little
difference in collapse, as the numerics confirm by the convergence of the
two curves. The most remarkable feature of the curves is their decline at
high frequencies. If $C(\scrn)$ has no lower bound, then in light 
of the scale invariance of the collapse this 
suggests that a very long wave train of very low amplitude news may in fact
be gravitationally unstable to collapse. Unsystematic experiments have 
produced data which collapse at $\BDV(\scrn)$ as low as 0.06, but
numerical experiments alone cannot give a definitive answer. This 
complements analysis in \cite{BDV} suggesting that for extremely slowly
decaying news data $C(\scrn)$ may go to zero.
   
\begin{figure}[t]
\centerline{\epsfig{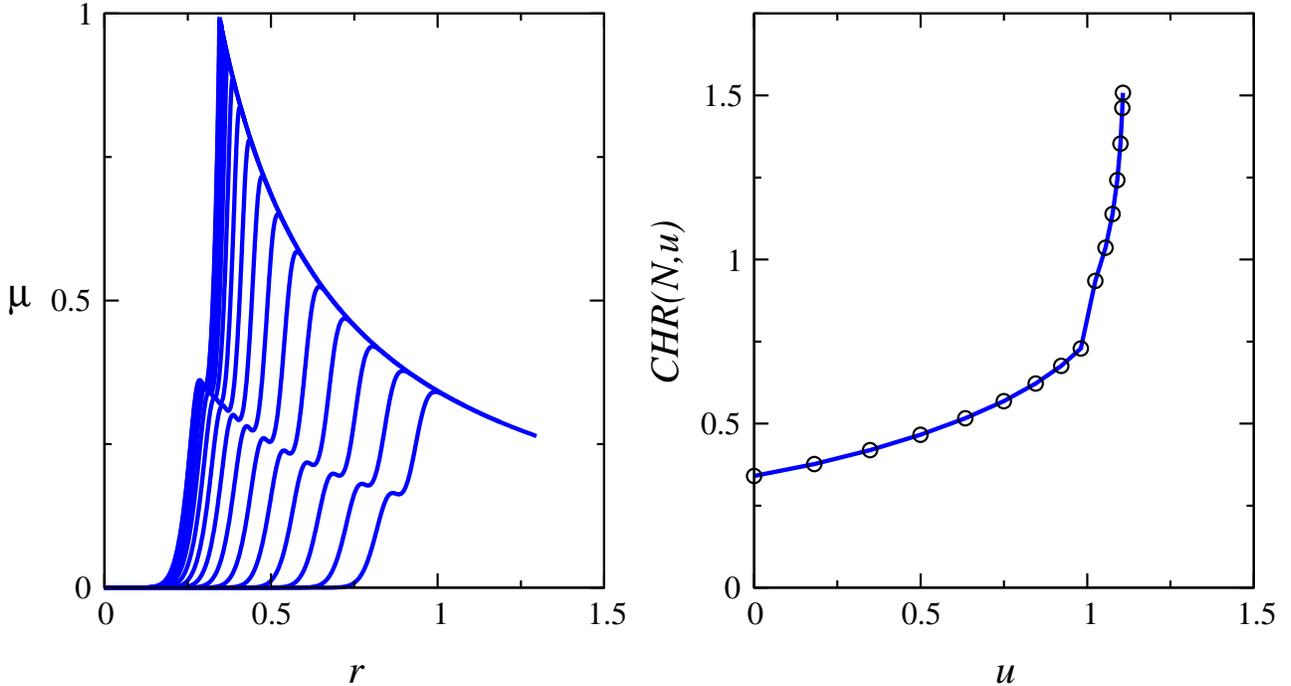}}
\caption{\em\baselineskip=11pt 
Collapse with $\scrn \propto n'(v)$, $f=0$, $w=1$ with amplitude
normalized to 0.4. Left: the evolution
of $\mu$ at several successive values of $u$ approaching horizon
formation. Right: The values of $\CHR(\scrn,u)$ for these values of $u$.}  
\label{chr4}
\end{figure}

\begin{figure}[t]
\centerline{\epsfig{file=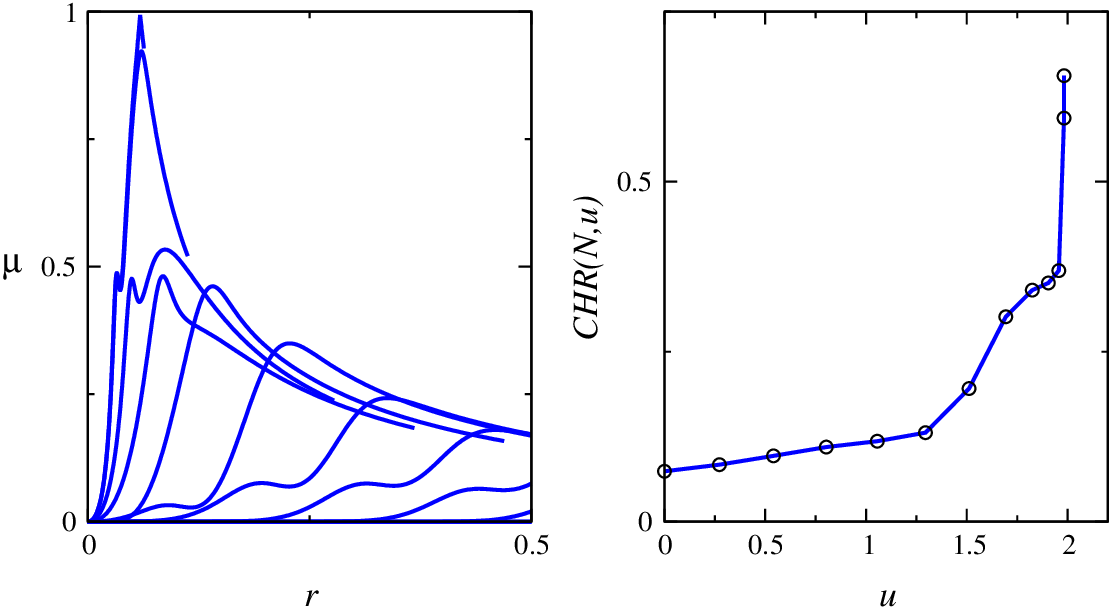,width=0.6\textwidth,angle=-90}}
\caption{\em\baselineskip=11pt 
Collapse with $\scrn \propto n'(v)$, $f=0$, $w=1$ and amplitude
normalized to 0.2. Left: the evolution
of $\mu$ at several successive values of $u$ approaching horizon
formation. Right: The values of $\CHR(\scrn,u)$ for these values of $u$.}  
\label{chr2}
\end{figure}

We also wish to make some comparison of this criteria with a rigorous
sufficient criterion of Christodoulou \cite{Chr3}. It states that if
for some value of $u$
\beq
\frac{2 \Delta m}{\Delta r }  \geq 
\left [ \frac{r_1}{r_2}\,\log \left (\frac{r_1}{2 \Delta r} \right )
+ \frac{6r_1}{r_2} -1 \right ]\,,
\label{chriscrit}
\eeq
where $r_1 < r_2, r_2 \leq 3r_1/2, \Delta r = r_2 - r_1$ and
$\Delta m = m(u,r_2) - m(u,r_1)$, then future collapse is inevitable.
In physical terms it states that if the energy flux contained between
two spheres is larger than a function involving
the ratio of the two radii, then we can predict collapse. 

Defining $C_l(\scrn,u,r_1,r_2)$ to be the left hand side of this inequality
and $C_r(\scrn,u,r_1,r_2)$ to be the right and
\beq
\CHR(\scrn,u)=\sup_{r_1,r_2 \atop r_1 < r_2 \leq 3r_1/2} 
C_l(\scrn,u,r_1,r_2)/C_r(\scrn,u,r_1,r_2),
\eeq
we can then state the criterion as simply that $\CHR(\scrn,u) \geq 1$ is 
sufficient to insure collapse.

As this 
criterion needs to be applied at finite $u$ (since $m \rightarrow 0$ 
as $u \rightarrow -\infty$, $\CHR$ can never predict 
collapse at large negative $u$) and the $\BDV$ criterion
needs to be applied at \scrim, a direct comparison
is impossible.
The best we can do to investigate its strength in predicting collapse
is to ask how early it can predict collapse before the formation
of the apparent horizon. 

To test this we form initial data $\scrn \propto n'(v)$ for 
$f=0$, $w=1$ corresponding to a gaussian pulse in $\phi$. Since the
normalized amplitude for threshold collapse of this data set
is found to be 0.1896, we first
normalize the amplitude of $\scrn(v)$ to 0.40 (Fig.~\ref{chr4})
to investigate
a heavy collapse and then to 0.20 (Fig.~\ref{chr2}) 
to investigate a near critical
collapse. In the first graph in each figure we show the profile
of $\mu$ versus $r$ for a number of values of $u$ leading up
to horizon formation. In the second graph we show the 
corresponding values of $\CHR(\scrn,u)$.
In the first case we find that even for the heavier
collapse the criterion predicts the appearance of the apparent horizon
only rather near to its formation. In the second we do not get near
enough to the horizon formation for the criterion to ever predict
the collapse. This would seem to point to the strong nonlinearities
involved in the horizon formation, particularly in the near critical
region where a very small change in the news amplitude leads to a
very large change in the largest value of $\mu$ eventually 
attained.

\section{Black Hole Interior Asymptotics and the PBB} 
\label{secasym}
\setcounter{equation}{0}

The proposal that `the PBB is as generic as gravitational collapse' is
the central proposal of \cite{BDV}, and the measure of validity 
of this proposal lies in the
interior of the black holes produced by massless scalar collapse. Since inside
the black holes the $r$ coordinate becomes timelike, the geometry there
describes a spacelike hypersurface approaching a future singularity and the
geometry is naturally interpreted as cosmological. And since the dilaton, 
the scalar field, also generically shows runaway behavior, this suggests
that the black hole interior when interpreted in the string frame may 
generically exhibit the pole-law inflation of the PBB scenario. With a 
graceful exit mechanism (possibly 
due to corrections to the lowest
order perturbative action coming from large curvature and strong coupling)
to regulate the approach to the singularity, the black hole interior
may be regarded as a `baby universe'. The first step is to check that 
the regime where the dynamics is dominated only by the lowest order action
without corrections has some resemblance to the PBB scenario. 

Thanks to the work of \cite{BK} it is known that the interior solutions for 
generic initial data (without additional field content) 
will be asymptotically velocity dominated, 
a `quiescent cosmological singularity'. The 
final cosmological approach to the singularity will take the form of
Kasner evolution, an FRW evolution modified by some degree of anisotropy.
As the PBB form of inflation is known not to wash out anisotropy \cite{KD}
(as is clear in our case, since once we are into the Kasner 
regime the degree of anisotropy freezes and remains constant),
unless we take the opinion that the graceful exit will 
also cure this problem, we should look for features of generic gravitational
collapse that will also select for isotropy. Working in spherical
symmetry takes us part of the way to isotropy. since this forces two of the
Kasner exponents to be equal, but this restriction is quite artificial. 
Nonetheless, if we can show collapse favors a third exponent equalling the
other two, this would be some evidence a non-spherical collapse could also 
favor isotropy.

In addition to relative isotropy, a `good' PBB collapse should
satisfy several requirements related to APT.
The dilaton should be increasing with 
$\Phi<<0$ and the spacetime curvature should be small at 
the entrance to the asymptotic regime
to ensure the lowest order action is valid for a sufficiently 
large number of e-folds of dilaton driven inflation. 
But we observe that any solution can be mapped to a solution
of arbitrarily small curvature and weak coupling
by applying a constant shift of the 
dilaton $\Phi \rightarrow \Phi+C$ or performing a scale transformation
$\Phi(r) \rightarrow \Phi(k r)$ (changing the scalar curvature $R \rightarrow
R/k^2$) without affecting the asymptotic Kasner behavior, as will be
seen from the following analysis. 

In this scenario the need for a `tuning' of these two classical moduli of the
collapse is acknowledged. But due to the invariances of the system
we should also expect collapses to occur on all size scales
and couplings. Only sufficiently
large and weakly coupled collapses will lead to universes such as we
occupy. Others are regarded as stillborn \cite{VenHou}. This also makes
the question of the degree of fine-tuning subject
to anthropic considerations, since the birth of a universe is no longer
a unique event. With these invariances, the task left to 
a numeric code will be to analyze the relative degree of isotropy
in the cosmology within the collapse as a function of the shape and
amplitude of the initial news data. 

We first recall the small $r$ asymptotics derived 
in \cite{BDV,Bur} (and confirmed numerically in \cite{Bur}) 
and rederive the relation between the asymptotics of the
different functions. We start with the dominant terms in the 
growth of the scalar as we approach the $r=0$ singularity,
\beq
\Phi(v,r) \sim 4 \sqrt{\pi} \hb \sim 2 \alpha(v) \log r + C_{\phi}(v).
\label{phiasym}
\eeq
Using the defining relation for $h$ (\ref{hdef}) we easily have,
\beq
h(v,r)=\hb(v,r)+{\alpha(v) \over {2 \sqrt{\pi}}}.
\eeq
Inserting this into (\ref{greq}) we find that to leading order,
\beq
g(v,r) &\sim& C_g(v) r^{\alpha^2(v)} \label{gasym} \\
\gb(v,r) &\sim& {{C_{\gb}(v)} \over r}.
\eeq
Recalling the relation $\gb/g=1-2 m/r$, we have in the limit of 
small $r$,
\beq
m(v,r) \sim C_m(v) r^{-\alpha^2(v)}.
\label{masym}
\eeq
So we see that nature of the evolution near the singularity is 
controlled entirely by $\alpha(v)$. We will next see how this $\alpha(v)$ is 
related to the Kasner exponents in the string and Einstein frame. We will
follow the discussion of \cite{BDV}, filling in some details and omitting
some more general discussions.

Inserting the limiting behavior above into (\ref{nullmet}) we find
\beq
ds^2 \sim r^{\alpha^2(v)-1} ~dv^2+r^{\alpha^2(v)} ~dv ~dr+r^2 d\omega^2.
\eeq
Here we have omitted all $C(v)$ factors (in the asymptotic Kasner region
they are locally nearly constant) and numerical constants since they 
can finally be absorbed into rescalings of the coordinates. By completing
the square, this can be written as
\beq
ds^2 \sim -{1 \over 4} r^{\alpha^2(v)+1} dr^2 + 
 (dv+{r \over 2} dr)^2 r^{\alpha^2(v)-1}+
 r^2 d\omega^2.
\eeq 
Changing to Einstein frame cosmic time via
\beq
(-t_E) \sim r^{{\alpha^2(v)+3} \over 2},
\eeq
this becomes (again omitting constants)
\beq
d{s_E}^2 &\sim& -d{t_E}^2+(-t_E)^{2 \lambda_1} d{x_1}^2+(-t_E)^{2 \lambda_2}
(d{x_2}^2+d{x_3}^2) \\
\Phi &\sim& \gamma(v) \log(-t_E),
\eeq
where $x_2$ and $x_3$ are the spacelike coordinates of spherical 
symmetry (in $d\omega^2$) and $x_1$ is the remaining spacelike 
direction and
\beq
\label{lam}
\lambda_1 (v) = \frac{\alpha^2 (v) - 1}{\alpha^2 (v) + 3} \,,
\quad \quad 
\lambda_2 (v) =  \frac{2}{\alpha^2 (v) + 3} \,,
\quad \quad 
\gamma (v) = \frac{4 \alpha (v)}{\alpha^2 (v) + 3} \,.
\eeq

We complete the transition into the string frame by applying (\ref{strtoein})
and changing to string frame cosmic time by
\beq
-t_S=(-t_E)^{{2+\gamma(v)} \over 2}
\eeq
obtaining
\beq
d{s_S}^2 &\sim& -d{t_S}^2+(-t_S)^{2 \alpha_1} d{x_1}^2+(-t_S)^{2 \alpha_2}
(d{x_2}^2+d{x_3}^2) \\
\Phi &\sim& \sigma(v) \log(-t_S)
\label{stringframe},
\eeq
where
\beq
\label{alp}
\alpha_1 (v) &=& \frac{\alpha^2 (v) +2 \alpha(v)- 1}
  {\alpha^2 (v) + 2 \alpha(v)+ 3}, \quad 
\alpha_2 (v) =  \frac{2 (\alpha(v)+1)}{\alpha^2 (v) + 2 \alpha(v)+ 3},
\nonumber \\
\sigma (v) &=& \frac{4 \alpha (v)}{\alpha^2 (v) + 2 \alpha(v)+ 3}.
\eeq

We can now state the requirement for a `good' PBB collapse in terms
of $\alpha(v)$. Consistent with
inflationary behavior, we should at least 
require all directions are expanding (all string Kasner exponents negative)
and an increasing dilaton. 
These requirements can be met by looking for regions where 
$-1-\sqrt{2}<\alpha(v)<-1$ with 
$\alpha(v)=-\sqrt{3}$ corresponding to exact isotropy. 
Since changing the sign of $\scrn(v)$ reverses the sign of $\alpha(v)$,
we will be concerned with the magnitude of $\alpha(v)$.

While it is clearly difficult to generalize about the infinite dimensional
space of all possible initial data, we will present several data sets
extracted from the many we have run
looking for general trends. We choose a collapsing data set and extract
values of $\alpha^2(v)$ by fits to the small $r$ data from 
(\ref{phiasym}), (\ref{gasym}) and (\ref{masym}). We will take agreement
between these values to indicate that we are qualitatively in the 
asymptotic regime.

\begin{figure}[t]
\centerline{\epsfig{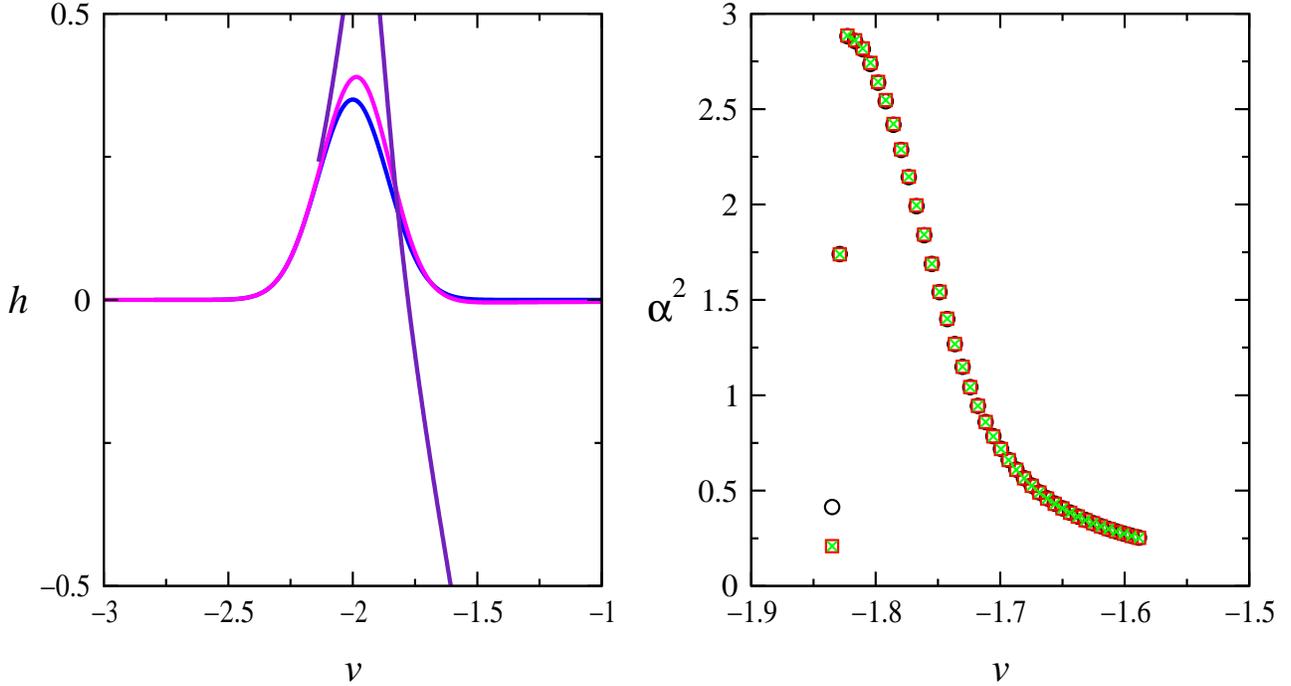}}
\caption{\em\baselineskip=11pt 
Collapse of an amplitude 0.35 gaussian news pulse, 
left: $h$ as a function of $v$ at
\scrim, the end of \scrim evolution and near horizon formation (in order of
increasing amplitude). Right: the extracted values of $\alpha^2(v)$
from $\phi$ ($\circ$), $m$ ($\Box$) and $g$ ($\times$).}
\label{alg3}
\end{figure}

\begin{figure}[t]
\centerline{\epsfig{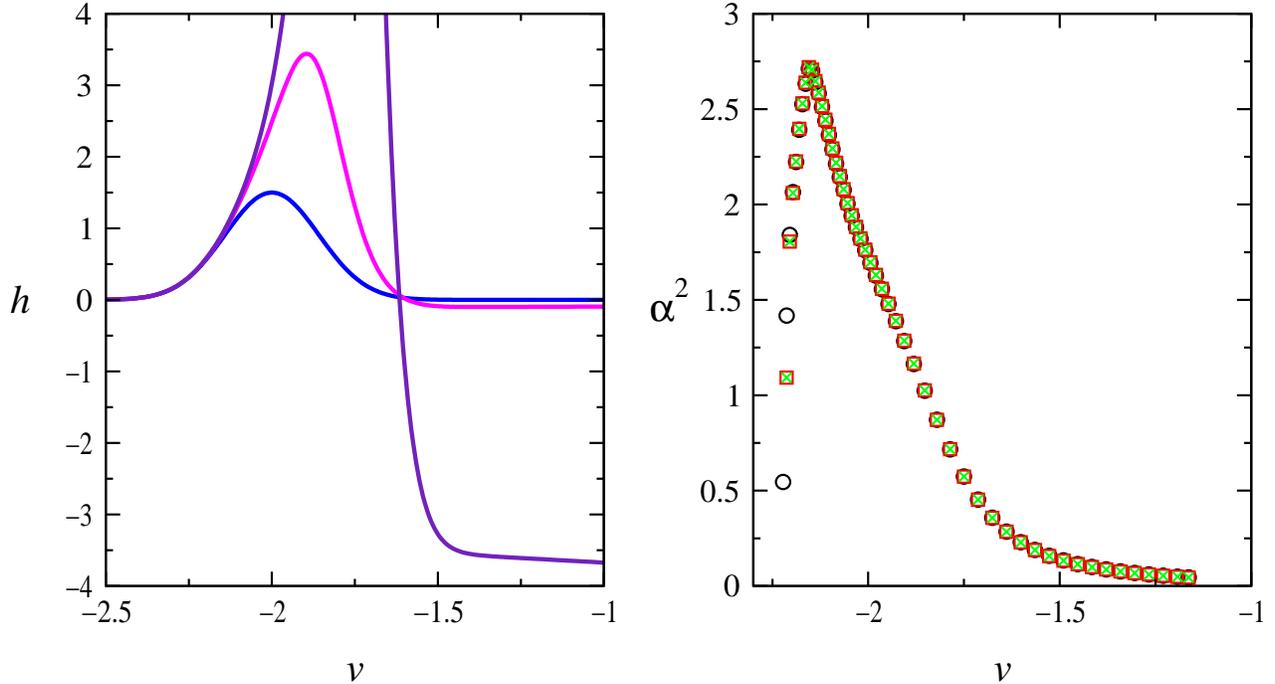}}
\caption{\em\baselineskip=11pt 
Collapse of an amplitude 1.5 gaussian news pulse, 
left: $h$ as a function of $v$ at
\scrim, the end of \scrim evolution and near horizon formation (in order of
increasing amplitude). Right: the extracted values of $\alpha^2(v)$
from $\phi$ ($\circ$), $m$ ($\Box$) and $g$ ($\times$).}
\label{alg15}
\end{figure}

In the first set, Fig.~\ref{alg3}, we choose gaussian news, 
$a=0.35$, $f=0$, $v_0=-2$ and $w=0.2$ (\ref{wavelet}). This 
is somewhat over the critical amplitude of $a_0=0.309$.
The first graph shows the profile of $h$ at \scrim, at the 
end of the \scrim evolution and at the point where $\mu>0.99$ and 
it shortly after forms a black hole of radius $r_{BH}=0.1349$. We run
the $v$-mode integration over the region near the $r=0$ 
singularity to extract the
$\alpha^2(v)$ values from data going as deep as $r_{min}=0.00008$. 
As the second figure shows, this includes the region of significant
deviation of $\alpha(v)$ from zero. At larger values of $v$, $\alpha(v)$ is
running to its Schwarzschild value of zero, in accordance with the 
dictum `black holes have no hair inside or out'. We do find
a region of $\alpha^2(v)>1$, corresponding a minimal inflationary 
criterion of `all-directions expanding', and further the maximum value of 
$\alpha^2(v) \sim 3$ reached corresponds to relative isotropy,
although it is a narrow region very near to the initial $v$ of the black
hole formation. The innermost points represent ingoing nulls that fail to 
penetrate into the asymptotic region, so the apparent fall-off in the 
value of $\alpha(v)$ should not be taken seriously.
We also note that the onset of black 
hole formation occurs as the end of the pulse is striking
the origin.

\begin{figure}[t]
\centerline{\epsfig{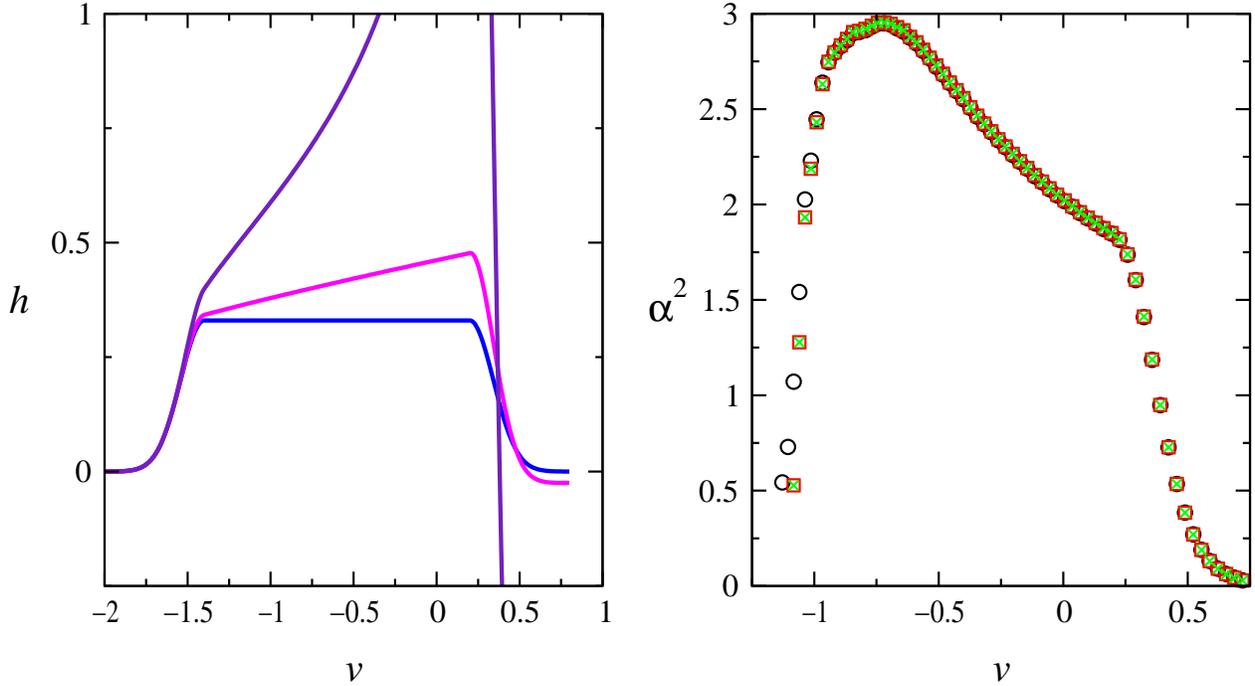}}
\caption{\em\baselineskip=11pt 
Collapse of a lengthened gaussian pulse, left: $h$ as a function of $v$ at
\scrim, the end of \scrim evolution and near horizon formation (in order of
increasing amplitude). Right: the extracted values of $\alpha^2(v)$
from $\phi$ ($\circ$), $m$ ($\Box$) and $g$ ($\times$).}
\label{mesa2}
\end{figure}

In the second set, Fig.~\ref{alg15} we raise the amplitude to $a=1.5$,
forming a black hole of radius $r_{BH}=3.495$. We cut off at
$r_{min}=0.002$. The qualitative results are quite similar to the 
first case, pointing again to a small relatively isotropic 
region corresponding to in a narrow region close to the onset of black
hole formation. With the increased amplitude this now occurs when
the leading surface of the pulse arrives at the origin.

In the third set, Fig.~\ref{mesa2} we extend a gaussian pulse 
of width 0.2 by adding a flat 
portion of $v$ length 1.6 in the
center of the pulse. We also tune the amplitude to 0.33 so the 
apparent horizon forms as the flat part of the pulse is striking
the origin, although the wave form is no longer flat by then. This
time the closeness of the approach to $\alpha^2=3$ is striking. 

Finally, in the fourth set we send in additional pulses after the 
first. We also vary the exact form of the initial news data to
\beq
h(v)=a \sin^2({{\pi (v-v_0)} \over w})
\eeq
with $v_0=-2$, $a=0.4$ and $w=1$, and setting $h=0$ outside of 
four peaks of the news function. Again we find an initial approach
to quasi-isotropy and notice that further incoming news data after
the formation of the black hole does not have much effect on 
interior asymptotics.

Comparison of these data with our criteria for a `good' initial PBB 
leads to limited good news. Large regions of the black hole interior
do not reach strong enough gradients in $\Phi$ to meet the `all directions
expanding' criterion, much less isotropy. 
On the other hand there is a strong hint
that there may be generically 
a small region very close the initial point of singularity
formation where these requirements are met. 

Indeed we note that
as we approach smaller and smaller $r$, closer and closer to the
first appearance of the apparent horizon at the origin, 
the asymptotic $\alpha^2(v)$
seems to approach 3 and seems to resist exceeding this value.
This tempts us to conjecture that singularity
onsets in spherical symmetry may approach exact isotropy. This is
also suggested by the fact the regular $r=0$ origin is a point
of exact isotropy before the collapse, 
so this quality may extend into the singular region.
While suggestive, we have been unable to verify this analytically.
However, this would imply the occurrence of this near isotropic region
may come from the exact spherical symmetry imposed, so it would
not clearly extend to nonspherical collapse.

\begin{figure}[t]
\centerline{\epsfig{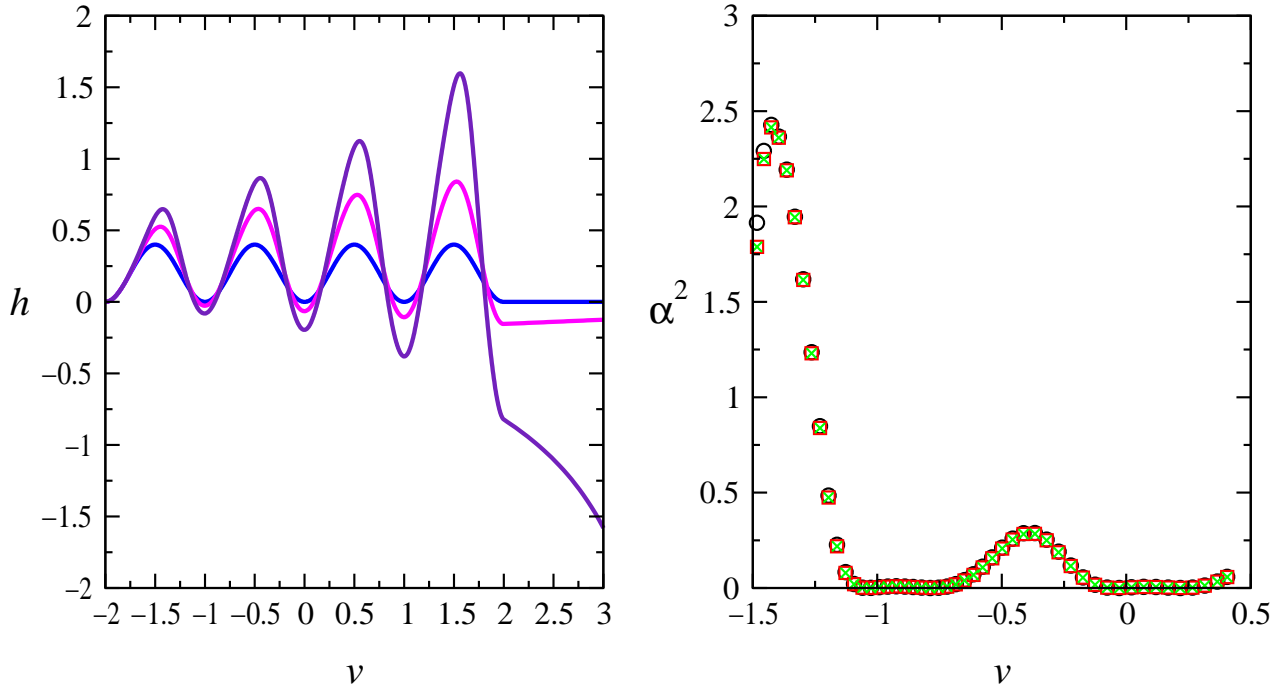}}
\caption{\em\baselineskip=11pt 
Collapse of a wave train, left: $h$ as a function of $v$ at
\scrim, the end of \scrim evolution and near horizon formation (in order of
increasing amplitude). Right: the extracted values of $\alpha^2(v)$
from $\phi$ ($\circ$), $m$ ($\Box$) and $g$ ($\times$).}
\label{4sine4}
\end{figure}

This singularity onset region is also the most 
numerically difficult to reach and
our results so far are hardly conclusive. Analyzing the exact 
solution (\cite{BDV} (4.37)) we find it corresponds to 
$\alpha^2(v)=1$. But this solution is neither asymptotically flat
nor differentiable at singularity onset, so it may not be representative
of more physical collapse data. There also seems to be some trend for 
near critical collapses to reach higher levels of isotropy, but it is
not clear whether this is because the isotropic regions are absent
for heavier collapse or that we are simply unable to reach them. 

While we conclude the string frame near-isotropy region seems to be 
generically small (though in a way difficult to quantify exactly),
it does seem to be present. Further, 
as suggested in \cite{Viso}, an initially small isotropic region may grow
preferentially relative to its more anisotropic neighbors. 
Along a slice of constant string frame time (where $t_S=0$ is the
time of singularity, lacking a graceful exit), the volume of 
space per unit $dv$ goes as 
\beq
Vol(t_S,v)=(-t_S)^{\Sigma \alpha_i(v)}.
\eeq
This expansion rate is indeed maximized at $\alpha(v)=-\sqrt{3}$ with
$\Sigma \alpha_i(v)=-\sqrt{3}$. Since successful PBB inflation is
expected to be accompanied by a large growth in $\Phi$, in turn 
giving many e-folds in string time and in turn of the scale 
factor, this would seem to say that even an initially small region
of isotropy could be exponentially enhanced over its neighbors having
less negative exponents of growth rate.

However, this process of comparing growth rates at different places
at the same `time' requires singling out a particular foliation of 
spacelike hypersurfaces. For example, computing the 
string frame volume growth rates along
slices of constant $t_E$ or constant $r$
results in a preference for $\alpha(v)$ with
relatively strong anisotropy. Given this ambiguity we need to invoke
some physical reason for a given choice of foliation.
This choice should have something to do with the details of the 
final graceful exit mechanism but it is not clear to us how to 
make this more precise without choosing some family of observers
to give the proper weight to different regions. Perhaps, finally, these
observers should correspond to the locations of future civilizations
who could observe this anisotropy. For a Bayesian look at this 
possibility we refer again to \cite{BDV}, since it is not clear
to us that the numerics can add quantitatively to this discussion.

Nevertheless, we routinely find collapses featuring regions
of $\alpha(v)<-1$ and indeed approaching isotropy, confirming
the general picture of `ballooning patches of inflating space'
in \cite{BDV}. 

\section{Conclusions} 
\label{conc}
\setcounter{equation}{0}

We have numerically investigated massless scalar collapse in spherical
symmetry in the context of its possible role as a precursor for a 
PBB inflationary cosmology. We find evidence that such collapses
do provide plausible precursors. Our most interesting conclusion based
on this preliminary survey, that 
there seems to be a region of substantial isotropy very near the 
initial appearance of the singularity on the central axis, 
will need to be investigated
further. It lies in a region that is numerically difficult to reach
and it would be much more satisfying to have some analytical understanding
of this phenomenon, especially the degree to which it is particular
to spherical symmetry. This conclusion also will need to be qualified
to the degree to which the additional field content of string theory
can spoil this behavior.

We have also investigated a proposed collapse criterion \cite{BDV}.
We find suggestions, again at the fringes of our numerical reach, that
special forms of news data, specifically long trains of low 
amplitude waves, can collapse even though the criterion would suggest
they should remain perturbative. In an extreme form this phenomenon 
would suggest that perhaps a very long train of {\it arbitrarily} low
amplitude may be capable of collapse through a long slow buildup of
non-linearity. While this seems somewhat physically unlikely it
would be interesting to try to get some sort of analytical handle 
on this sort of behavior as well. 

\section{Acknowledgements} 
\label{ack}
\setcounter{equation}{0}

The author wishes to thank Gabriele Veneziano for his continued
support of this project and his valuable suggestions and the CERN Theory
Division for their hospitality. Also thanks for useful talks and
comments from Alessandra Buonanno, Thibault Damour, Vincent Montcrief, 
Phillipos Papadopoulus, I. M. Segal, Reza Tavakol and Carlo Ungarelli. 
We also express gratitude 
for computational facilities provided by the UMS CNRS/Polytechnique
MEDICIS and CERN.

\end{document}